\documentclass[conference, letterpaper]{IEEEtran}
\ifCLASSINFOpdf
\else
\fi
\ifCLASSINFOpdf
   \usepackage[pdftex]{graphicx}
\else
\fi

\usepackage[cmex10]{amsmath}
\usepackage{color}
\usepackage{fancyhdr}
\usepackage[caption=false,font=footnotesize]{subfig}

\renewcommand{\thispagestyle}[2]{}

\fancypagestyle{plain}{
        \fancyhead{}
        \fancyhead[C]{first page center header}
        \fancyfoot{}
        \fancyfoot[C]{first page center footer}
}
\usepackage{ctable,multirow,amsmath,graphicx}
\usepackage{url, comment}

\DeclareMathOperator*{\argmax}{arg\,max}

\begin{document}

%
\title{Speaker Change Detection Using Features through A Neural Network Speaker Classifier}

\author{\IEEEauthorblockN{Zhenhao Ge, Ananth N. Iyer, Srinath Cheluvaraja, Aravind Ganapathiraju}
\IEEEauthorblockA{Interactive Intelligence Inc., Indianapolis, Indiana, USA\\
Email: \{roger.ge, ananth.iyer, srinath.cheluvaraja, aravind.ganapathiraju\}@inin.com}
}


\maketitle

\begin{abstract}
The mechanism proposed here is for real-time speaker change detection in conversations, which firstly trains a neural network text-independent speaker classifier using in-domain speaker data. Through the network, features of conversational speech from out-of-domain speakers are then converted into likelihood vectors, i.e. similarity scores comparing to the in-domain speakers. These transformed features demonstrate very distinctive patterns, which facilitates differentiating speakers and enable speaker change detection with some straight-forward distance metrics. The speaker classifier and the speaker change detector are trained/tested using speech of the first 200 (in-domain) and the remaining 126 (out-of-domain) male speakers in TIMIT respectively. For the speaker classification, 100\% accuracy at a 200 speaker size is achieved on any testing file, given the speech duration is at least 0.97 seconds. For the speaker change detection using speaker classification outputs, performance based on 0.5, 1, and 2 seconds of inspection intervals were evaluated in terms of error rate and F1 score, using synthesized data by concatenating speech from various speakers. It captures close to 97\% of the changes by comparing the current second of speech with the previous second, which is very competitive among literature using other methods.  
\end{abstract}



%
\IEEEpeerreviewmaketitle

\section{Introduction}
\label{sec:intro}

Speaker Change Detection (SCD) is a task to detect the change of speakers during conversations. An efficient and accurate speaker change detector can be used to partition conversations into homogeneous segments, where only one speaker is presented. Speaker recognition or verification can then be performed on the clustered speaker segments, rather than on a frame-by-frame basis, to improve accuracy and reduce cost. However, SCD is challenging since prior information of the speakers is absent, and it is usually required to detect speaker change in real-time, within limit delay, e.g. within 1 or 2 seconds of speech. 

SCD can be divided into retrospective vs. real-time detection \cite{nielsen2013efficient}. The former one is normally based on model training for speakers and detection algorithm, using Gaussian Mixture Models (GMMs) and Hidden Markov Models (HMMs), etc. \cite{anguera2012speaker}. It includes approaches with different thresholding criteria, such as Bayesian Information Criterion (BIC) \cite{wooters2008icsi}, Kullback-Leibler (KL)-based metrics \cite{rougui2006fast}, etc. For the real-time detection, the decision has to be made using limited preceding data with low computational cost. Research has been focused on improving features and developing efficient distance metrics. Lu et al. \cite{lu2002speaker} obtained reliable change detection in real-time news broadcasting with the Bayesian feature fusion method. In the evaluation using TIMIT synthesized data by Kotti et al. \cite{kotti2006automatic}, it mean F1 score was $0.72$ and it observed a significant drop in accuracy for speaker change within durations less than 2 seconds. Another work from Ajmera et al. \cite{ajmera2004robust} reported 81\% recall and 22\% precision using BIC and log-likelihood ratios on HUB-4-1997 3-hour news data. 

Here a novel real-time mechanism for SCD is presented. It first transforms conversations to speaker classification outputs through a feed-forward neural network, trained by in-domain speaker data; then detects if speaker change is presented by comparing the similarity of adjacent intervals of 0.5, 1, or 2 seconds. Though the speakers presented in the testing conversations are usually out-of-domain (unseen) speakers to the network, and the outputs merely serve as the likelihood of them to be classified to the in-domain speakers, the pattern of the new speakers is still revealed in the network outputs and can be used to distinguish one another. This enables us to develop some straight-forward distance metrics to capture speaker change. Very promising performance is achieved on the synthesized conversational speech using TIMIT, which is not feasible if we directly use the raw features without going through the speaker classification network.

\begin{figure*}[htb]
 \centering
  \begin{tabular}{c}
 	\includegraphics[height=3.25cm, width=10.5cm]{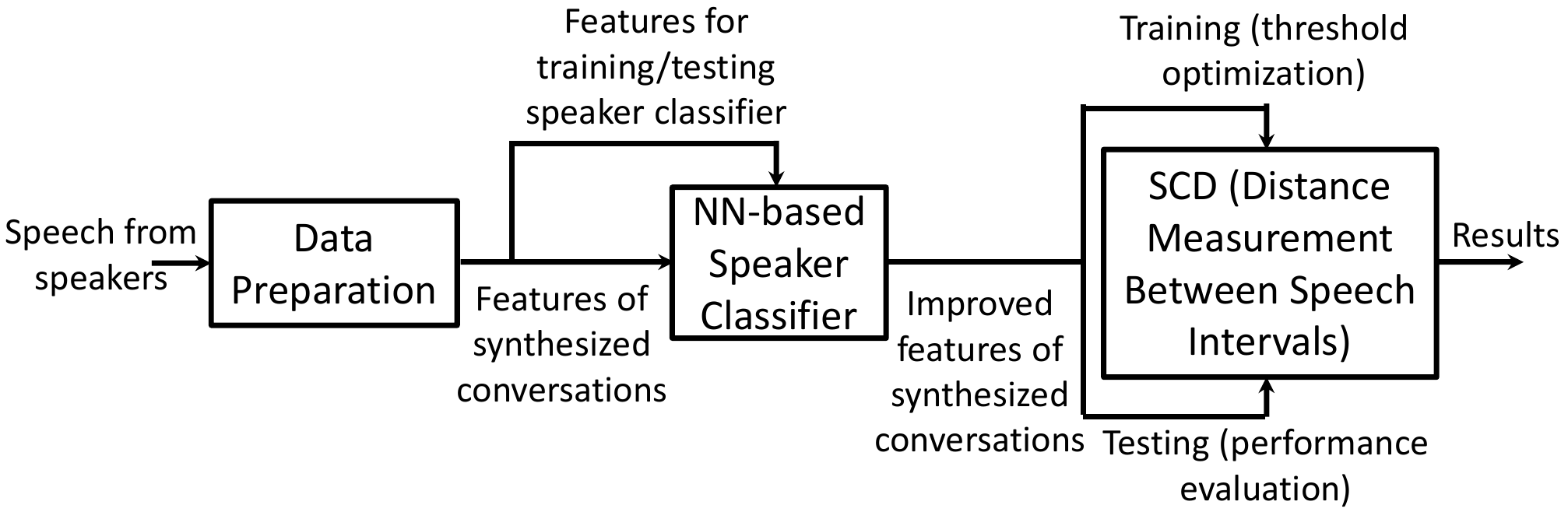}
  \end{tabular}
  \caption{Diagram of using improved features through an NN-based speaker classifier for Speaker Change Detection (SCD). \label{fig:diagram}}
\end{figure*}

Fig. \ref{fig:diagram} shows a global picture for using an NN-based speaker classifier as a feature transformer and then detecting speaker changes with the improved features.
The following sections walk through 3 major components in Fig. \ref{fig:diagram}, including data preparation, (Sec. \ref{sec:data}), the framework of neural network (NN) based speaker classification (Sec. \ref{sec:nnsc}) and the speaker change detection mechanism, i.e. the distance metrics that we use for detection based on speaker classification outputs (Sec.  \ref{sec:nnscd}). Finally, the conclusion and future work is in Sec. \ref{sec:conclusion}. 

\section{Data Preparation}
\label{sec:data}

Speech of all 326 male speakers in the ``train'' folder of the TIMIT corpus is used here. Data of males from the ``test'' folder and data of females from both ``train'' and ``test'' folders are currently not used. For each speaker, there are 10 data files containing one sentence each from 3 categories: ``SX'' (5 sentences), ``SI'' (3 sentences) and ``SA'' (2 sentences). The 326 male speakers are sorted alphabetically and divided into 2 groups: first 200 speakers (group A) and remaining 126 speakers (group B).
For group A, sentences in the ``SX'' and ``SI'' categories are different among speakers. They are combined with a total duration around $20$ seconds per speaker and used to train the text-independent neural network speaker classifier. Sentences in the ``SA'' category are the same and shared with all speakers, so they can be used to test the accuracy with no distinguishable information added through content. For group B, synthesized conversations are generated by concatenating speech from multiple speakers. Conversations created using ``SX'' and ``SI'' sentences of the first 63 out of 126 speakers are used to find the optimal threshold to determine speaker change, while conversations with ``SX'' and ``SI'' sentences of the remaining 63 speakers are used for testing the SCD performance. 

The following 2 subsections introduce the process of converting raw speech into features used in the development of the speaker classifier and SCD algorithm, including a) preprocessing, and b) feature extraction and concatenation.  

\subsection{Preprocessing}
\label{subsec:preprocessing}

Preprocessing mainly consists of a) scaling the maximum of absolute amplitude to 1, and b) Voice Activity Detection (VAD) to eliminate the unvoiced part of speech. Experiments show both speaker classification and speaker change detection can perform significantly better if speakers are evaluated only using voiced speech, especially when the data is noisy.

An improved version of Giannakopoulos's recipe \cite{giannakopoulos2009method} with short-term energy and spectral centroid is developed for VAD. Given a short-term signal $s(n)$ with $N$ samples, the energy is:
\begin{equation}
\label{eq:ste}
E = \frac{1}{N} \sum_{n=1}^{N}|s(n)|^2 , 
\end{equation}
and given the corresponding Discrete Fourier Transform (DFT) $S(k)$ of $s(n)$ with $K$ frequency components, the spectral centroid can be formulated as:
\begin{equation}
\label{eq:spectral centroid}
C = \frac{\sum_{k=1}^{K}kS(k)}{\sum_{k=1}^{K}S(k)} .
\end{equation}
The short-term energy $E$ is used to discriminate silence with environmental noise, and the spectral centroid $C$ can be used to remove non-environmental noise, i.e. non-speech sound, such as coughing, mouse clicking and keyboard tapping, since they normally have different spectral centroids compared to human speech. Only when $E$ and $C$ are both above their thresholds $T_{E}$ and $T_{C}$, the speech frame is considered to be voiced, otherwise, it will be removed. These thresholds are adjusted to be slightly higher to enforce a stricter VAD algorithm and ensure the quality of the captured voiced sections. This is achieved by tuning the signal median smoothing parameters, such as step size and smoothing order, as well as setting the thresholds $T_{E}$ and $T_{C}$ as a weighted average of the local maxima in the distribution histograms of the short-term energy and spectral centroid respectively. In this work, the TIMIT speech with the original $16$K sampling rate is segmented into overlapped frames with a $50$ ms window size and a $25$ ms hop size. 

\subsection{Feature Extraction, Normalization and Concatenation}
\label{subsec:feature}

\begin{figure}[htb]
 \centering
  \begin{tabular}{c}
 	\includegraphics[height=3cm, width=6cm]{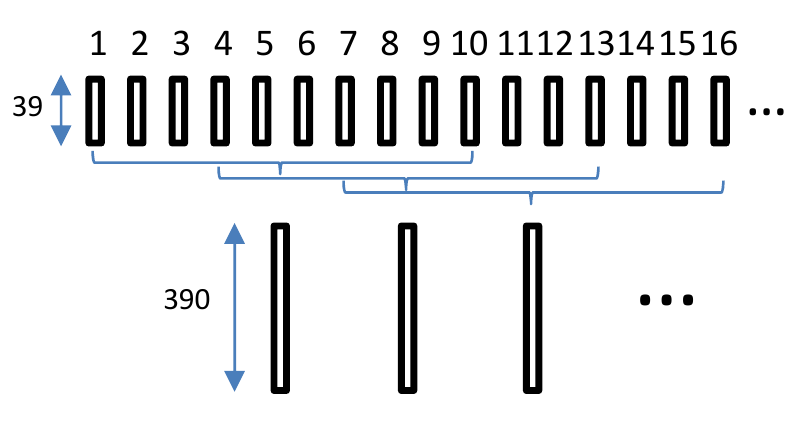}
  \end{tabular}
  \caption{Feature concatentation example with a window size of 10 frames and a hop size of 3 frames. \label{fig:feature_concatenation}}
\end{figure}

The 39-dimensional Mel-Frequency Cepstral Coefficients (MFCCs) with delta and double delta were generated from the preprocessed speech, following Ellis's recipe \cite{Ellis05-rastamat}. They were extracted using overlapped $25$ ms Hamming windows which hop every $10$ ms. Then, the features of each speaker were normalized with his own mean and variance. To capture the transition patterns within longer durations, these 39-dimensional feature frames were concatenated to form overlapped longer frames. In this work,  10 frames ($100$ ms) were concatenated with hop size of 3 frames ($30$ ms) as shown in Fig. \ref{fig:feature_concatenation}.   

\section{Neural Network Speaker Classification}
\label{sec:nnsc}

The concatenated features (e.g. 390 dimensional feature vectors) are used as the input to a  neural network speaker classifier. As mentioned in the first paragraph of Sec. \ref{sec:data}, the ``SX'' and ``SI'' sentences of the first 200 male speakers were used for training, and the remaining ``SA'' sentences from the same set of speakers were used for testing.

\subsection{Cost Function and Model Structures}
\label{subsec:cost}

Ng's neural network training recipe for hand-written digit classification \cite{couseraml} is used here, which treats the multi-class problem as $K$ separate binary classifications. It is considered to be the generalization of the cost function of binary classification using logistic regression, which is built on slightly different concepts compared with the cross-entropy cost function with softmax as the output layer \cite{srihari}. 

Given $M$ samples, $K$ output classes, and $L$ layers, including input, output and all hidden layers in between, the cost function can be formulated as:
\begin{eqnarray}
\label{eq:cost_function}
J(\Theta) &=& -\frac{1}{M}  \left[ \sum_{m=1}^{M}\sum_{k=1}^{K} \left( y_{k}^{(m)} \log (h_{\theta}(x^{(m)})_{k}) \right. \right. \\ \nonumber 
    & &  + \left. \left. (1-y_{k}^{(m)}) \log (1-h_{\theta}(x^{(m)})_{k}) \right) \right] \\ \nonumber
    & & + \frac{\lambda}{2M}\sum_{l=1}^{L-1}\sum_{i=1}^{s_{l}}\sum_{j=1}^{s_{l+1}}(\theta_{ji}^{(l)})^{2}
\end{eqnarray}
where $h_{\theta}(x^{(m)})_{k}$ is the $k$th output of the final layer, given $m$th input sample $x^{(m)}$, and $y_{k}^{(m)}$ is its corresponding target label. The $2$nd half of Eq.  (\ref{eq:cost_function}) is the regularization factor to prevent over-fitting, where $\lambda$ is the regularization parameter and $\theta_{ji}^{(l)}$ is the $j$-th row, $i$-th column element of the weight matrix $\Theta^{(l)}$ between $l$-th and $(l+1)$-th layers, i.e. the weight from $i$-th node in $l$-th layer to $j$-th node in $(l+1)$-th layer. 

In this work, there is only 1 hidden layer ($L=3$) with $200$ nodes ($s_{2}=200$), the input feature dimension is $390$ ($s_{1}=390$), and the speaker classifier was trained with data from $200$ speakers ($s_{3}=K=200$). Therefore, the network structure is $390:200:200$, with weight matrices $\Theta^{(1)}$ ($200 \times 391$) and $\Theta^{2}$ ($200 \times 201$). The additional 1 column is a bias vector, which is left out in regularization, since the change of bias is unrelated to over-fitting. In this example, the regularization part in Eq. (\ref{eq:cost_function}) can be instantiated as
\begin{eqnarray}
\label{eq:cost_function_example}
\sum_{l=1}^{L-1}\sum_{i=1}^{s_{l}}\sum_{j=1}^{s_{l+1}}(\theta_{ji}^{(l)})^{2} = \
\sum_{i=1}^{390}\sum_{j=1}^{200}(\theta_{j,i}^{(1)})^2 \
     + \sum_{i=1}^{200}\sum_{j=1}^{200}(\theta_{j,i}^{(2)})^2.
\end{eqnarray}
%
 

\subsection{Model Training and Performance Evaluation}
\label{subsec:model}

The neural network model is trained through forward-backward propagation. Denoting $z^{(l)}$ and $a^{(l)}$ as the input and output of the $l$-th layer, the sigmoid function
\begin{equation}
\label{eq:sigmoid}
	a^{(l)} = g(z^{(l)}) = \frac{1}{1 + e^{-z^{(l)}}}  
\end{equation}
is selected as the activation function, and the input $z^{(l+1)}$ of the $(l+1)$-th layer can be transformed from the output $a^{(l)}$ of the $l$-th layer, using $z^{(l+1)} = \Theta a^{(l)}$. Then, $h_{\theta}(x)$ can be computed through forward propagation: $x = a^{(1)} \rightarrow z^{(2)} \rightarrow a^{(2)} \rightarrow \cdots \rightarrow z^{(L)} \rightarrow a^{(L)} = h_\theta(x)$. The weight matrix $\Theta^{(l)}$ is randomly initiated using continuous uniform distribution between $(-0.1, 0.1)$ and then trained through backward propagation of $\partial{J}/\partial{\theta_{j,i}^{(l)}}$, by minimizing $J(\Theta)$ using Rasmussen's conjugate gradient algorithm, which handles step size (learning rate) automatically with slope ratio method\cite{rasmussen2006gaussian}. 

\begin{figure}[htb]
\begin{minipage}[b]{1\linewidth}
  \centering
  \centerline{\includegraphics[height=4cm, width=6cm]{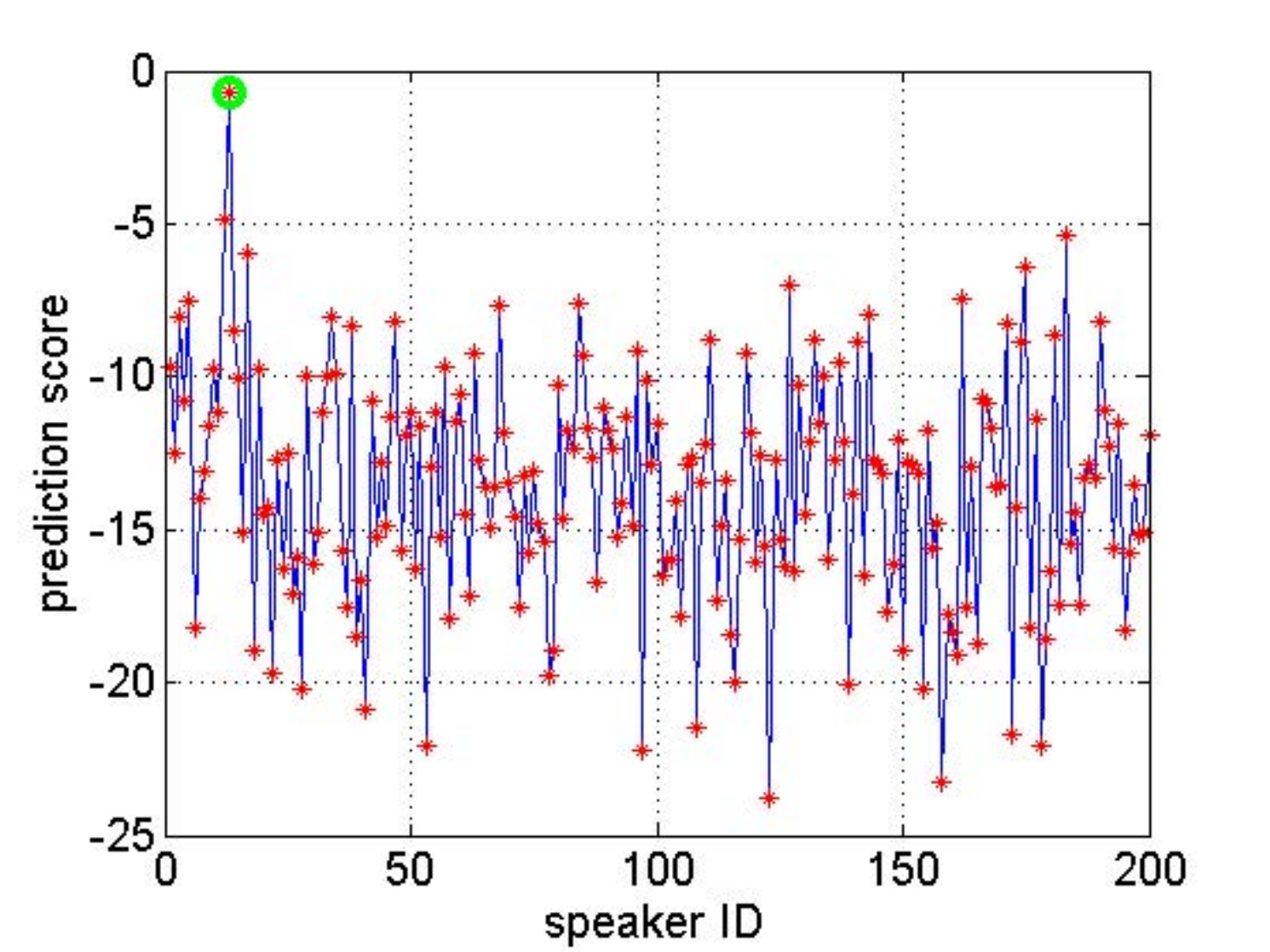}}
  \centerline{(a) {\tt SI1410} in training}\medskip
\end{minipage}
\hfill
\begin{minipage}[b]{1\linewidth}
  \centering
  \centerline{\includegraphics[height=4cm, width=6cm]{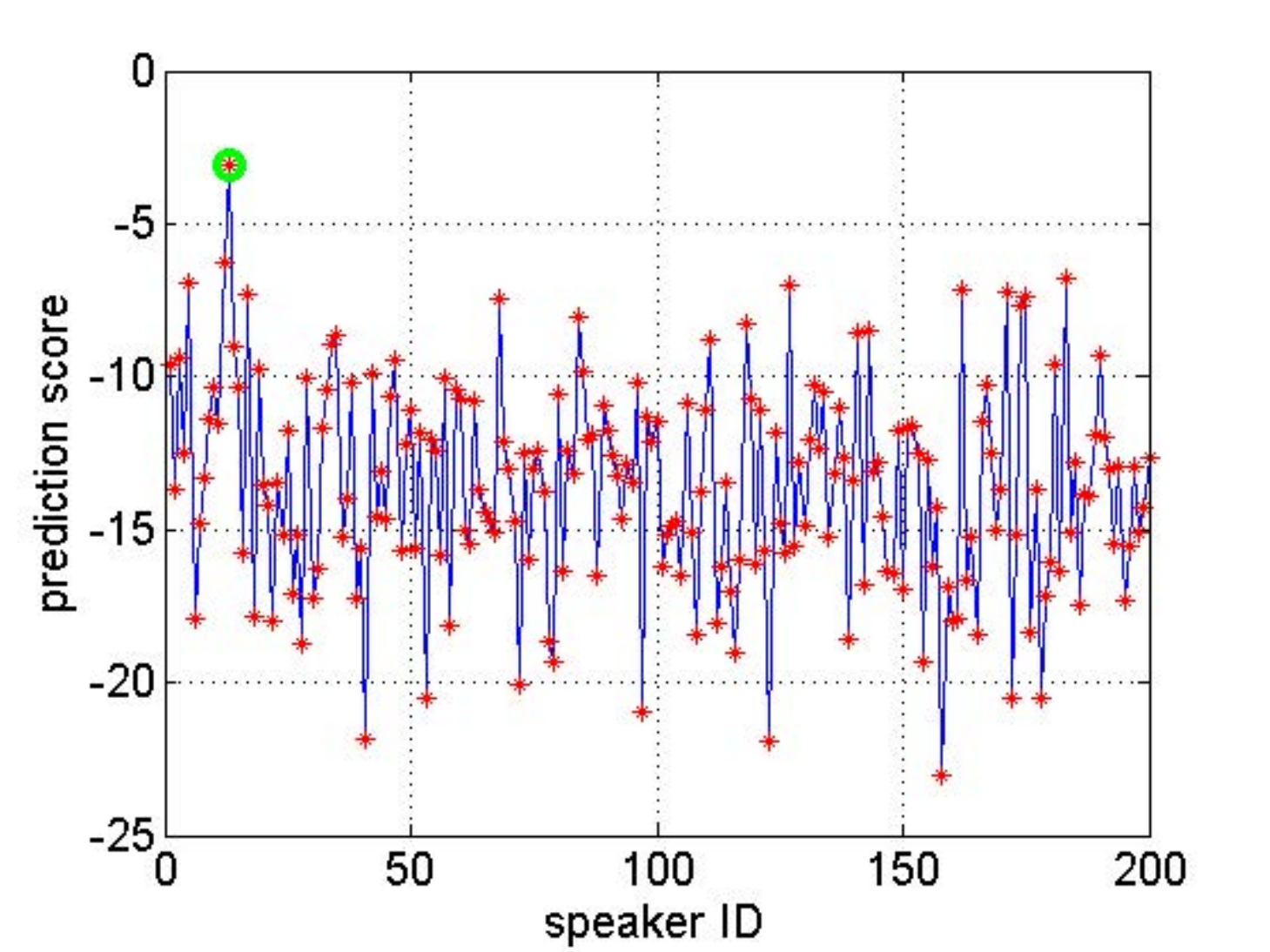}}
  \centerline{(b) {\tt SA1} in testing}\medskip
\end{minipage}
\caption{File-level prediction scores of $13$th speaker (MPGR0) in training and testing sets respectively.}
\label{fig:classifiation_example}
\end{figure}

In evaluating the classifier performance, the sigmoid output of the final layer $h_{\theta}(x^{(m)})$ is a $K$-dimensional vector, each element in the ranges of $(0,1)$. It serves as the ``likelihood'' to indicate how likely it is to classify $m$-th input frame into one of the $K$ speakers. The speaker classification can be predicted by the sum of log likelihood of $M$ input frames (prediction scores), and the predicted speaker ID $k^{*}$ is the index of its maximum:
\begin{equation}
\label{eq:prediction}
	k^{*} = \argmax_{k \in [1,K]} \left( \sum_{m=1}^{M} \log (h_{\theta}(x^{(m)})_{k}) \right) .
\end{equation}
$M$ can range from 1 to the entire frame length of the testing file. If $M=1$, the accuracy achieved is based on individual frames, each of which is $100$ ms (window duration $T_{win}$ in feature concatenation) with $30$ ms of new data, compared with the previous frame. On the other hand, if $M$ is equal to the total number of frames in file, the accuracy is file-based. The average duration of sentences (i.e. file length) is about 2.5 seconds. In general, larger $M$ leads to higher accuracy. Given the best model available with the network structure $390:200:200$, Fig. \ref{fig:classifiation_example} demonstrates an example of file-level prediction score of $13$-th speaker (MPGR0). It shows the peak of positives (in the green circle) is slightly dropped but still distinguishable enough to all other negatives, from the file {\tt SI1410} in the training set, to the file {\tt SA1} in the testing set.

Using this model, the file-level training and testing accuracies at $200$ speaker size are both 100\%, as indicated in Table \ref{tab:classification_accuracy}. 
\begin{table}[htb]
\footnotesize
  \caption{NN-based speaker classification performance with first 200 male in 16K TIMIT ($0.1$ sec./frame, $\sim$2.5 sec./file)}
  \label{tab:classification_accuracy}\centering
  \setlength{\tabcolsep}{2.25pt}
  \begin{tabular}{@{} *{6}{c} @{}} \toprule%
    \multirow{2}*{\textbf{Dataset}} &  \multicolumn{2}{c}{\textbf{Accuracy (\%)}} & \multicolumn{3}{c}{\textbf{Frames (seconds) needed for 100\% accuracy}} \\
     & \textbf{\textit{frame}} & \textbf{\textit{file}} & \textbf{\textit{min}} & \textbf{\textit{mean}} & \textbf{\textit{max}} \\ \midrule
     train & 96.63 & 100 & 2 (0.13) & 2.80 (0.15) & 6 (0.25) \\
     test & 79.65 & 100 & 5 (0.22) & 11.59 (0.42) & 30 (0.97) \\\bottomrule
  \end{tabular}
\end{table} 
The frame-level testing accuracy is $79.65$\%, which indicates that $79.65$\% frames in the testing set, with duration as little as $0.1$ second, can be classified correctly. It also shows the minimum, mean, and maximum number of consecutive frames needed and their corresponding durations in order to achieve 100\% accuracy, evaluated through all files in both training and testing datasets. Since the next frame provides only $30$ms (hop duration $T_{hop}$ in feature concatenation) additional information, compared with the current frame, given the number of frames needed $N$, the formula to compute the corresponding required duration $T$ is
\begin{eqnarray}
\label{eq:frame-duration}
	T = (N-1) \times T_{hop} + 1 \times T_{win} . 
\end{eqnarray}
With this formula, it requires only 11.59 frames (0.42 second) on average, to achieve 100\% accuracy in the testing dataset. 

Using the training data to test is normally not legitimate, and here it is used merely to get a sense of how the accuracy drops when switching from training data to testing data. 

\subsection{Model Parameter Optimization}
\label{subsec:optimization}

The current neural network model with the structure $390:200:200$ is actually the best one in terms of highest frame-level testing accuracy, after grid searching on a) the number of hidden layers ($1, 2$), and b) the number of nodes per hidden layer ($50, 100, 200, 400$), with a subset containing only 10\% randomly selected training and testing data. 

Once the ideal network structure is identified, the model training is conducted with a regularization parameter $\lambda$ in the cost function $J(\Theta)$, which is iteratively reduced from 3 to 0 through training. This dynamic regularization scheme is experimentally proved to avoid over-fitting and allow more iterations to reach a refined model with better performance.   


The training is set to be terminate once the testing frame accuracy cannot be improved more than $0.1\%$ in the last 2 consecutive training iterations, which normally takes around $500$ to $1000$ iterations. The training set is at $200$ speaker size with $20$ seconds speech each. It is fed in as a whole batch of data, which requires about 1 hour to train, on a computer with i7-3770 CPU and 16 GB memory. Therefore, the computational cost is certainly manageable.     

\section{Speaker Change Detection Using Speaker Classification Outputs}
\label{sec:nnscd}

\begin{figure}[htb]
\begin{minipage}[b]{1\linewidth}
  \centering
  \centerline{\includegraphics[height=4cm, width=6cm]{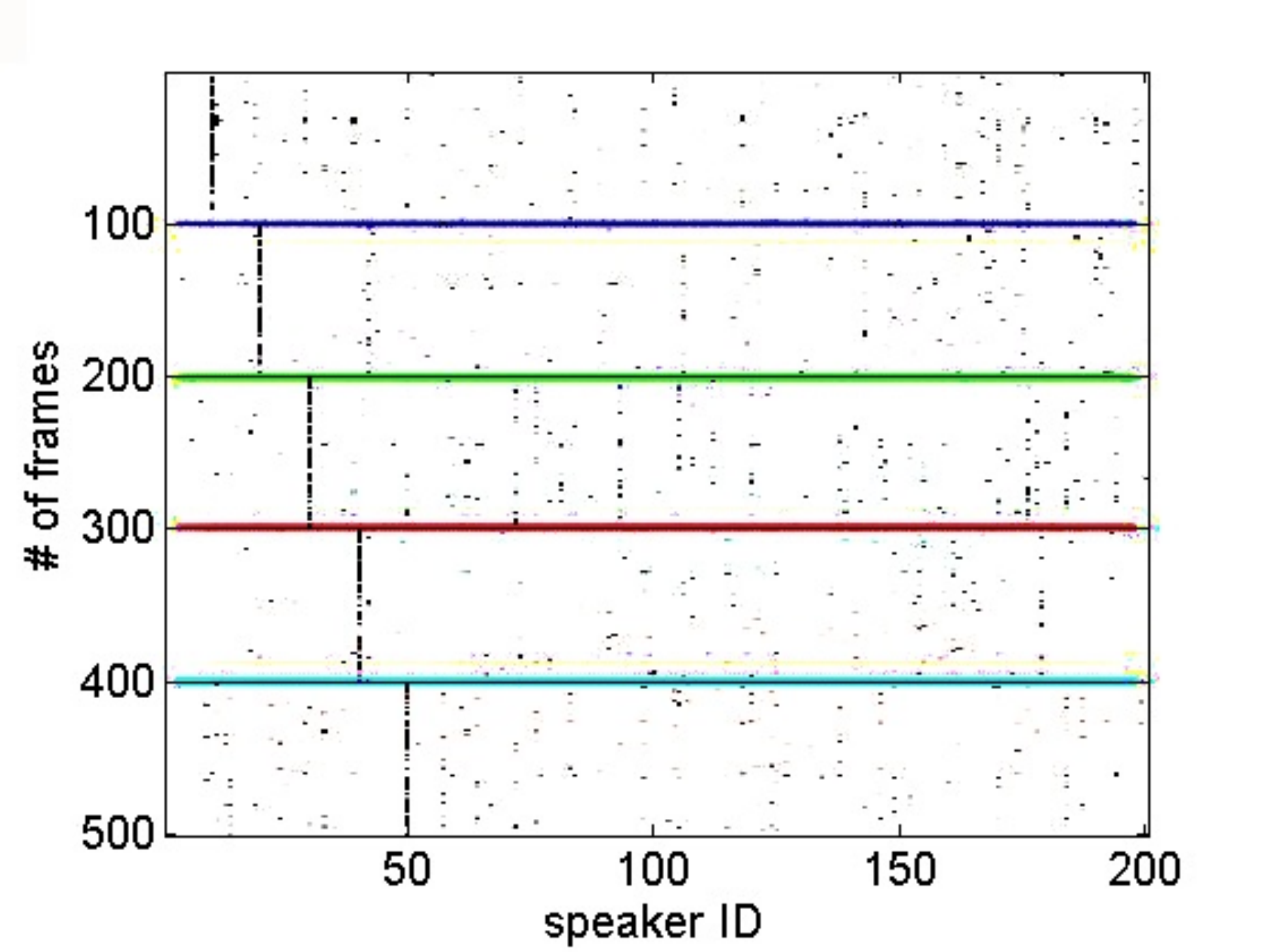}}
  \centerline{(a) 5 in-domain speakers}\medskip
\end{minipage}
\hfill
\begin{minipage}[b]{1\linewidth}
  \centering
  \centerline{\includegraphics[height=4cm, width=6cm]{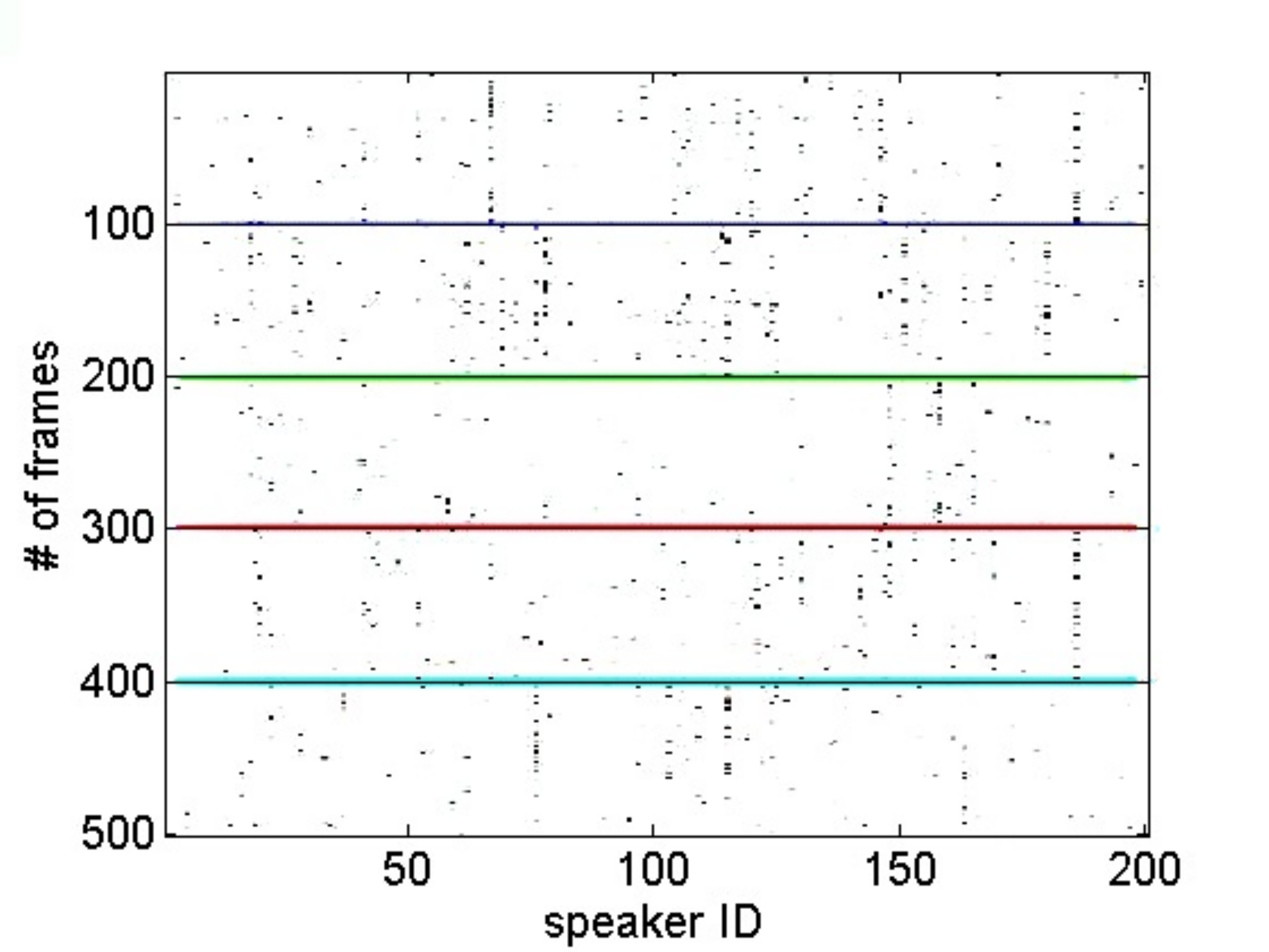}}
  \centerline{(b) 5 out-of-domain speakers}\medskip
\end{minipage}
\caption{Prediction output pattern visualization for in-domain and out-of-domain speakers.}
\label{fig:prediction_visualization}
\end{figure}

The main task in this work is to detect the speaker change in conversations. Developing an NN-based speaker classifier is among one of the approaches to improve features for that purpose. Here given the raw feature $x \in {\rm I\!R}^{390}$, the transformed new feature is denoted as
\begin{equation}
	d = \log (h_{\theta}(x)) \in {\rm I\!R}^{200} .
\end{equation} 
Dividing the conversation into consecutive speech intervals with equal frame length $M$, the goal is to develop some distance metrics to measure the difference between 2 sets of improved features at current interval $t$ and previous interval $t-1$, which is formulated as:
\begin{equation}
\label{eq:metrics}
d_{t}^{'} = \mathrm{dist}(d_{t}, d_{t-1})
\end{equation}
Fig. \ref{fig:prediction_visualization} shows an example of concatenation of 100 200-dimensional transformed features for 5 in-domain and 5 out-of-domain speakers. These features are reversed to linear scale (i.e. $h_{\theta}(x)$) rather than logarithmic scale for better visibility and are from the testing set containing ``SA'' sentences. The in-domain speakers are with speaker ID: 10, 20, 30, 40 and 50 (selected from first 200 speakers), while the IDs for the out-of-domain ones are: 210, 220, 230, 240 and 250 (selected from speakers with ID 201 to 326). The prediction scores are shown in gray scale, the larger the darker. The pattern for each speaker in (a) is fairly clear since they peak at their own ID indices. The pattern for speakers in (b) is not apparent, but one can still find some ``strip lines'', which indicate the consistency in similarity comparing one out-of-domain speaker with all in-domain speakers. 

\subsection{Distance Metrics to Compare Adjacent Intervals}
\label{subsec:metrics}

\begin{figure}[htb]
 \centering
  \begin{tabular}{c}
 	\includegraphics[width = .4\textwidth]{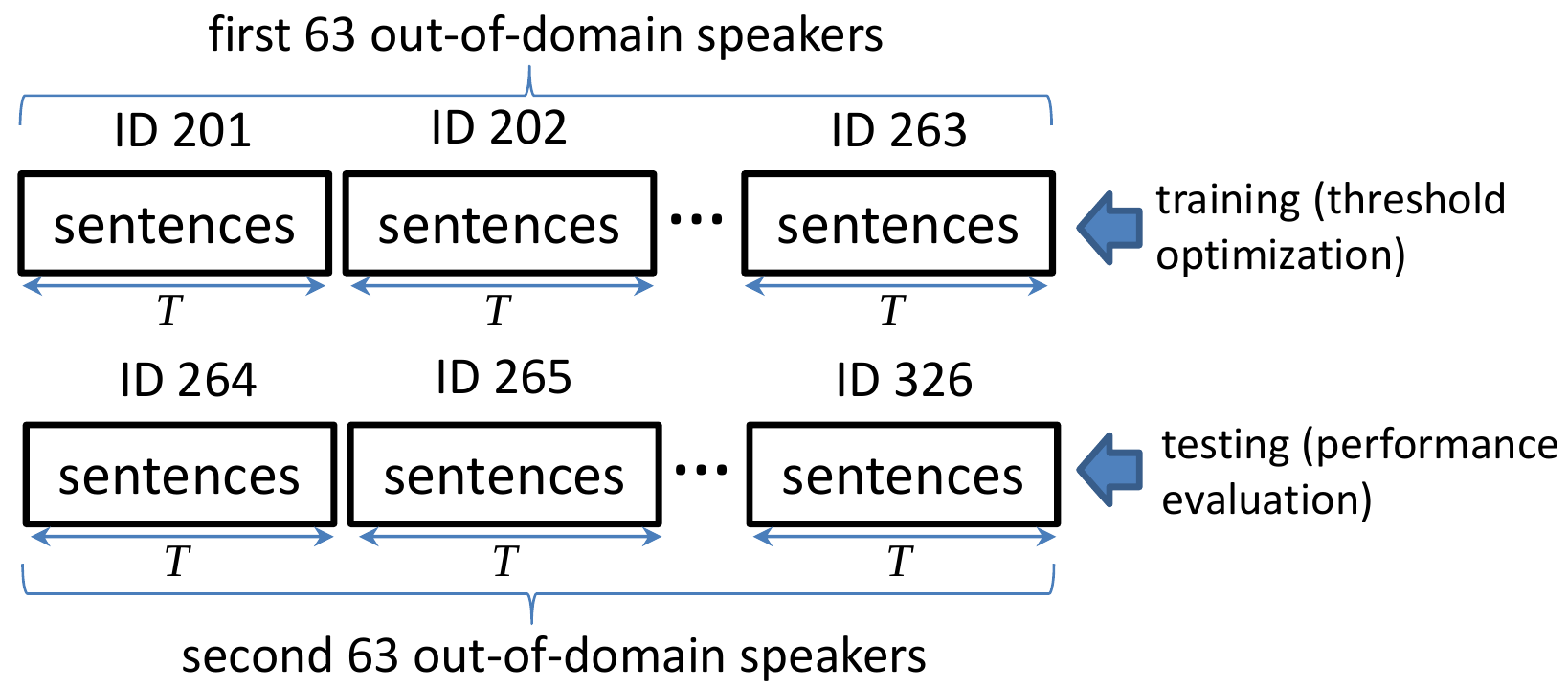}
  \end{tabular}
  \caption{Speech concatenation to form the synthesized conversations for training and tesitng in speaker change detection. \label{fig:speech_concatenation}}
\end{figure}

With the ``SX'' and ``SI'' sentences in the remaining 126 out-of-domain male speakers, 2 concatenated speeches are created using the data from the first 63 and the remaining 63 speakers respectively. They are used for training (threshold determination) and testing (performance evaluation) respectively in SCD. Sentences for the same speaker are first concatenated with the speech in the first $T$ seconds. $T$ is the duration for the shortest concatenation among all 126 speakers ($T = 14$ seconds in this work). These sentences grouped by speakers are then concatenated again to form the synthesized training and testing conversations ($14 \times 63 = 882$ seconds $\approx 15$ minutes each), as shown in Fig. \ref{fig:speech_concatenation}.   

The concatenated speech is then examined in each adjacent but non-overlapped interval $t$ with $M$ frames. Using $p$-norm distance metrics, Eq. (\ref{eq:metrics}) can be instantiated as:
\begin{equation}
\label{eq:p-norm}
	d_{t}^{'} = \left( \sum_{k=1}^{K} ( \vert \bar{d}_{t} - \bar{d}_{t-1} \vert ^{p} ) \right)^{\frac{1}{p}} ,
\end{equation}  
where $K$ is the number of in-domain speakers used to train the speaker classifier, i.e. dimension of the transformed features. $d_{t}$ and $d_{t-1}$ both are feature matrices at size of $M \times K$, and $\bar{d}_{t}$, $\bar{d}_{t-1}$ are their mean vectors with dimension $K$. The difference between current and previous intervals $d_{t}^{'}$ should be low (as negative), if feature matrices $d_{t}$ and $d_{t-1}$ belong to the same speaker, and should be high (as positive) vise versa. In this work, $p = \{\frac{1}{8}, \frac{1}{4}, \frac{1}{2}, 1, 2, 4, 8, \infty\}$ are tested, and $p = 2$, i.e. the Euclidean distance provided the best separation between positive (higher value expected) and negative (lower value expected) samples. 

Some other distance metrics other than $p$-norm, such as Bhattacharyya distance for comparison between 2 sets of samples, is also evaluated here. However, since the major difference between $\bar{d}_{t}$ and $\bar{d}_{t-1}$ demonstrated only with a few dimensions, which is much smaller than the full dimension $K$, the covariance matrices for both $\bar{d}_{t}$ and $\bar{d}_{t-1}$ are not positive definite, and this type of distance is not feasible then.

\subsection{SCD Training and Testing}
\label{subsec:scd_implementation}

\begin{figure*}[htb!]

\begin{minipage}[b]{.33\linewidth}
  \centering
  \centerline{\includegraphics[width=5.0cm]{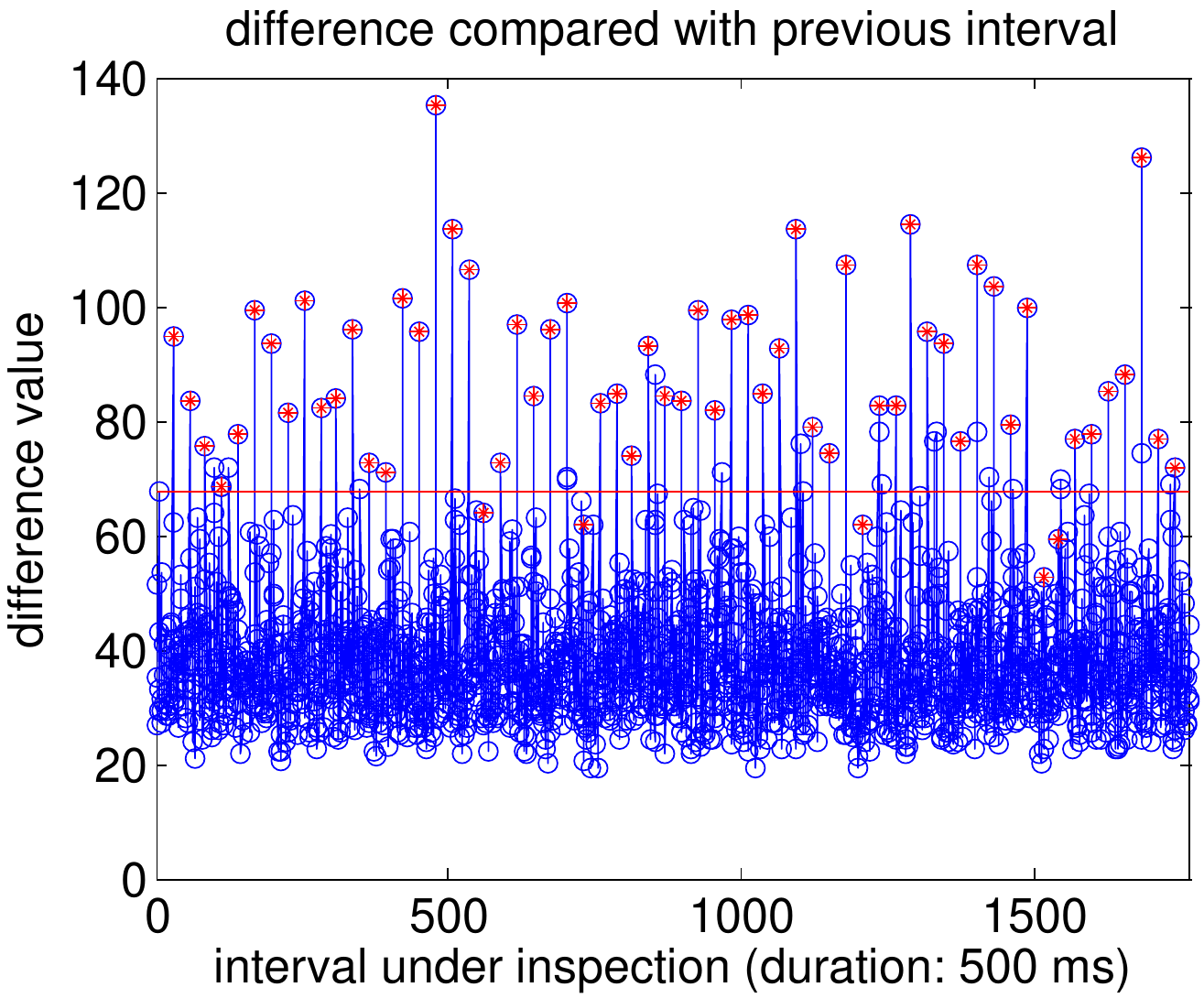}}
  \centerline{(a) experimental (0.5 sec.)}\medskip
\end{minipage}
\begin{minipage}[b]{0.33\linewidth}
  \centering
  \centerline{\includegraphics[width=5.0cm]{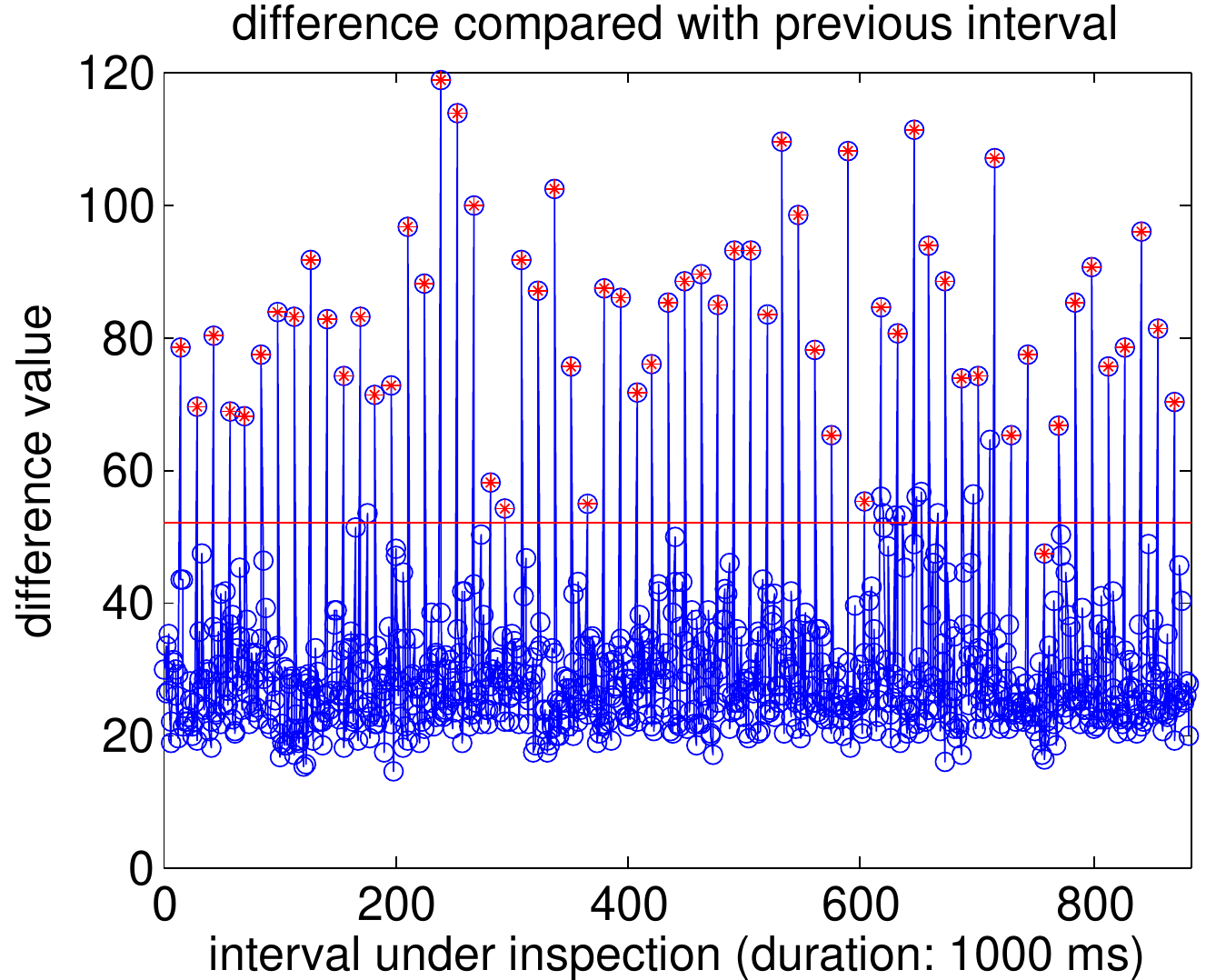}}
  \centerline{(b) experimental (1 sec.)}\medskip
\end{minipage}
\begin{minipage}[b]{.33\linewidth}
  \centering
  \centerline{\includegraphics[width=5.0cm]{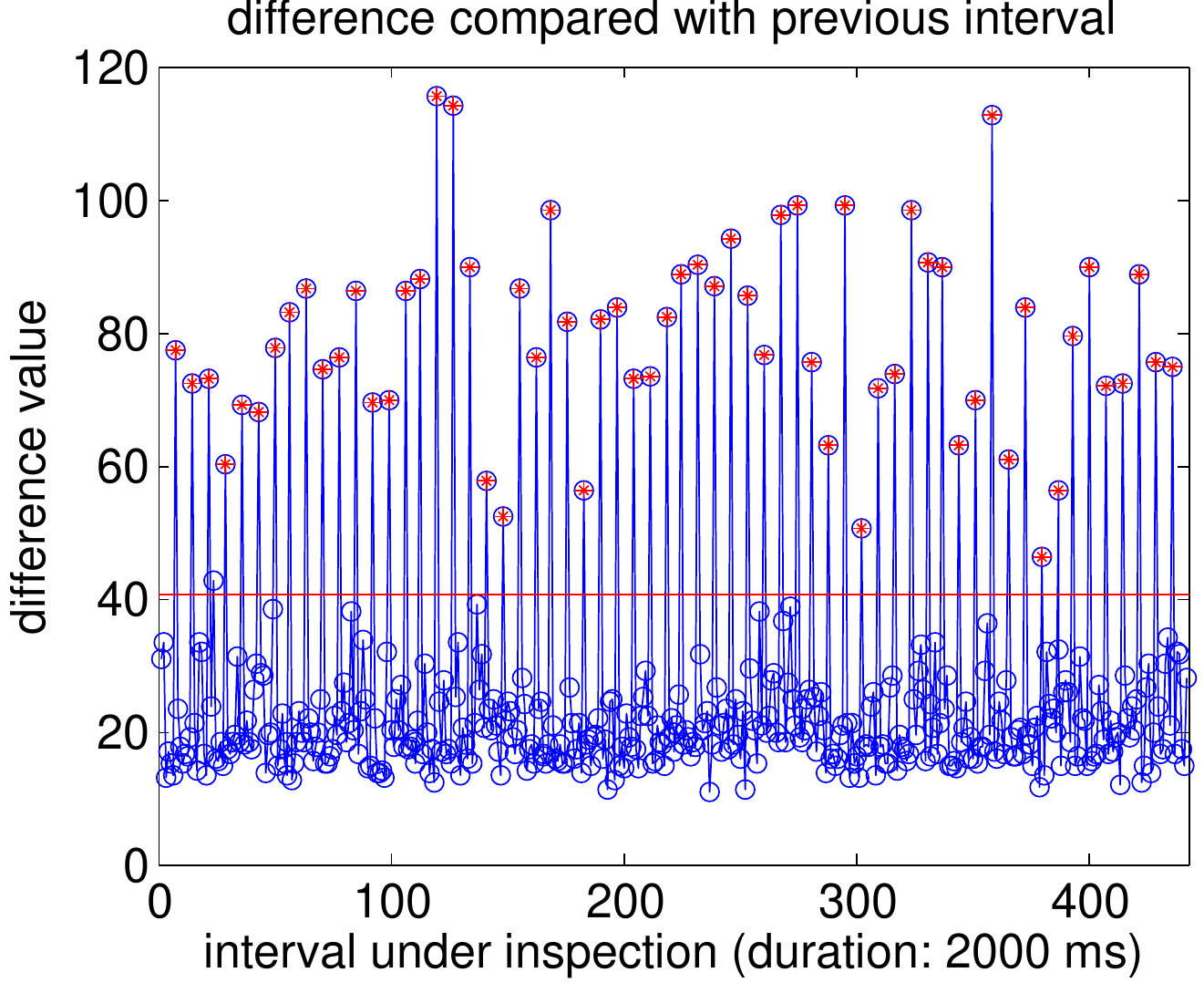}}
  \centerline{(c) experimental (2 sec.)}\medskip
\end{minipage}
\begin{minipage}[b]{0.33\linewidth}
  \centering
  \centerline{\includegraphics[width=5.0cm]{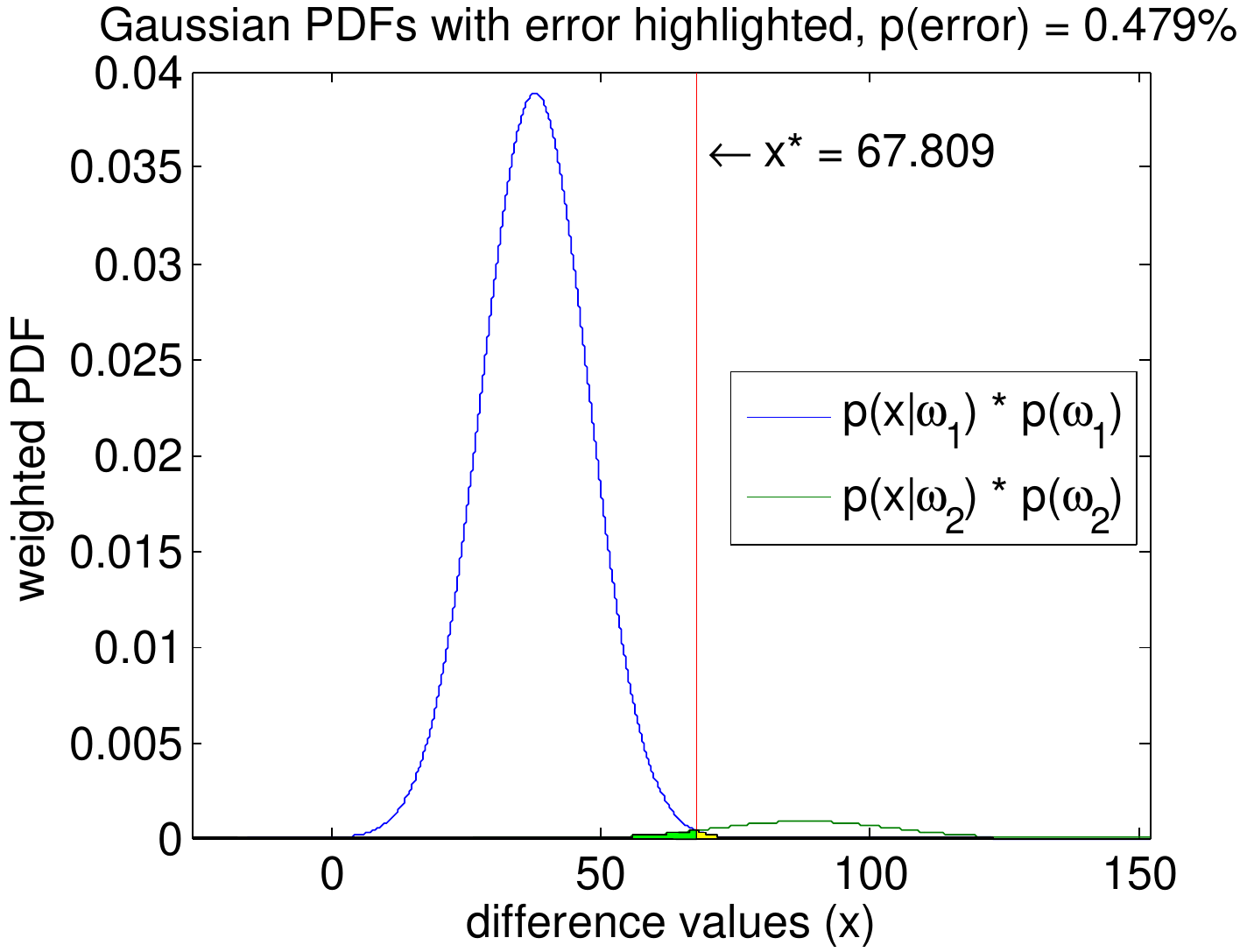}}
  \centerline{(d) theoretical (0.5 sec.)}\medskip
\end{minipage}
\begin{minipage}[b]{.33\linewidth}
  \centering
  \centerline{\includegraphics[width=5.0cm]{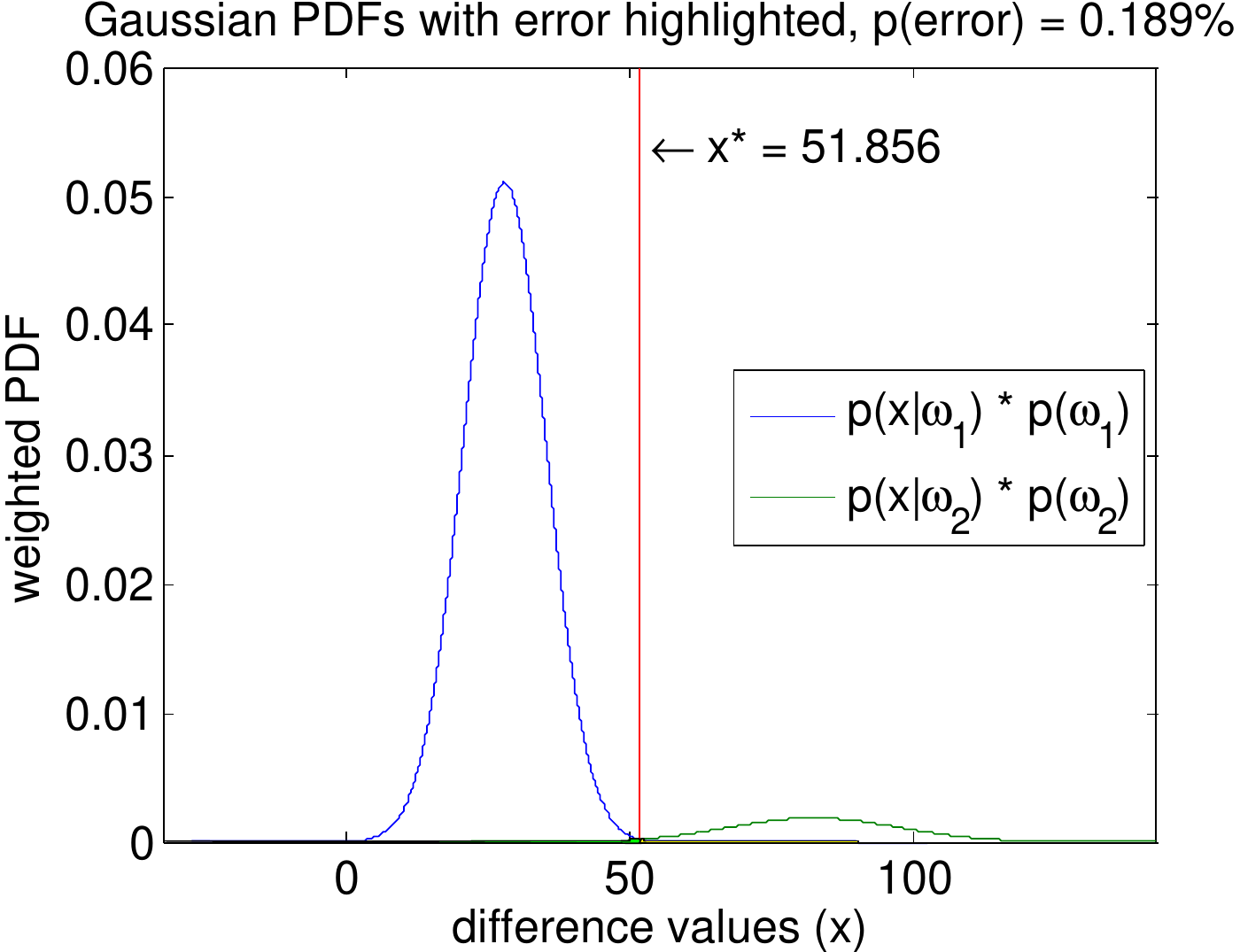}}
  \centerline{(e) theoretical (1 sec.)}\medskip
\end{minipage}
\begin{minipage}[b]{0.33\linewidth}
  \centering
  \centerline{\includegraphics[width=5.0cm]{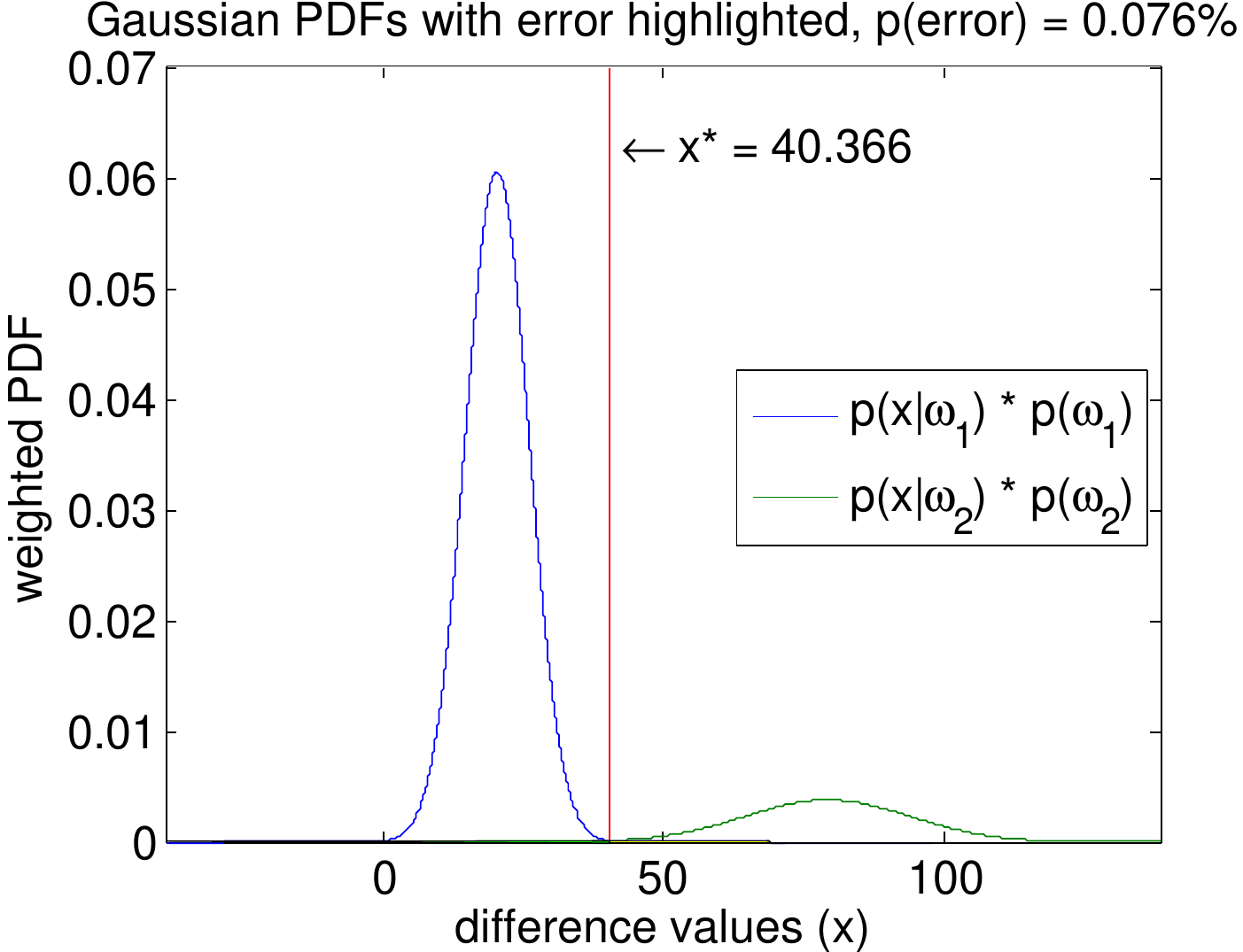}}
  \centerline{(f) theoretical (2 sec.)}\medskip
\end{minipage}
\caption{Experimental and theoretical distributions of positive and negative samples with multiple interval durations.}
\label{fig:scd_train}

\end{figure*}

Denoting the difference $d_{t}^{'}$ between current and previous intervals $t$, $t-1$ as sample $x$, the speaker changes can be detected if $x$ is higher than optimal threshold $x^{*}$. Fig. \ref{fig:scd_train} (a, b, c) plot $d_{t}^{'}$ vs. interval $t$ with interval durations 0.5, 1, and 2 seconds, where positive samples are highlighted with red stars. They are evenly distributed since the conversation speech is concatenated using speeches from individual speakers with same duration. By modeling the positive and negative samples as two Gaussian distributions, the Bayesian decision boundary is selected as the optimal threshold $x^{*}$. 

As is shown in Fig. \ref{fig:scd_train} (d, e, f), the negative samples (class label $\omega_{1}$ in Fig. \ref{fig:scd_train}) are much more than positive samples (class label $\omega_{2}$ in Fig. \ref{fig:scd_train}), especially when the time interval is small. Therefore, the dataset is very skewed. Therefore, F1 score is used along with error rate $P_{e}$ to measure the SCD performance. Given False Negative Ratio (FNR), i.e. ratio of classifying positive as negative ($FN$) vs. all positive ($P$), False Positive Ratio (FPR), i.e. ratio of classifying negative as positive ($FP$) vs. all negative ($N$) and $P = TP + FN$ and $N = TN + FP$, $P_{e}$ and $F1$ can be computed by:
%
%
\begin{equation}
\label{eq:error_rate}
P_{e} = \frac{FN + FP}{P+N}
\end{equation}
\begin{equation}
\label{eq:f1-score}
F1 =  \frac{2TP}{2TP+FP+FN}
\end{equation}
Table \ref{tab:scd_performance} show all these statistics for performance evaluation.

\begin{table}[htb]
\footnotesize%
  \caption{SCD performance on synthesized conversations (theoretical on the training set, experimental on the testing set), with multiple inspection interval.}
  \label{tab:scd_performance}\centering
  \setlength{\tabcolsep}{1.5pt}
  \begin{tabular}{@{} cc|cccc|cccc @{}} \toprule%
  	itvl./ & itvl. & $P_{e}$(\%) & F1 & FNR(\%) & FPR(\%) & $P_{e}$(\%) & F1 & FNR(\%) & FPR(\%) \\
  	spkr. & sec. & & \multicolumn{2}{c}{(theoretical)} & & & \multicolumn{2}{c}{(experimental)} &  \\ \midrule
	28 & 0.5 & 0.479 & 0.929 & 10.288 & 0.121 & 2.042 & 0.747 & 14.516 & 1.587 \\
	14 & 1 & 0.189 & 0.987 & 2.022 & 0.050 & 0.454 & 0.969 & 0 & 0.488 \\
	7 & 2 & 0.076 & 0.997 & 0.412 & 0.020 & 0.227 & 0.992 & 0 & 0.265 \\
     \bottomrule
  \end{tabular}
\end{table} 

The results for training data is theoretical, computed using Gaussian distributions in Fig. \ref{fig:scd_train} (d, e, f), and the ones for testing data is experimentally counted, using plots similar to Fig. \ref{fig:scd_train} (a, b, c). However, the optimal thresholds for the training data may not be still optimal for the testing data. It shows above $10\%$ of speaker changes cannot be detected by comparing features in the current and previous 0.5 second interval, i.e. FNR is $10.288\%$ or $14.516\%$ at theoretical and experimental cases. However, these numbers drop significantly once the inspection interval gets longer. 

\subsection{Potential Further Improvement}
\label{subsec:further_improvement}

The approach described above for SCD is checking the difference $d_{t}^{'}$ between $d_{t}$ and $d_{t-1}$, features in the current and previous intervals. However,  by comparing current difference $d_{t}^{'}$ with previous difference $d_{t-1}^{'}$ and the next difference $d_{t+1}^{'}$, i.e. difference of the difference, may reveal more reliable information. This is based on the assumption that if speaker change occurs in the current interval, the $d_{t}^{'}$ will be much higher than both its previous and next ones, $d_{t-1}^{'}$ and $d_{t+1}^{'}$. This distance metric can be considered as ``second derivative'' of the raw feature, and is formulated as:
\begin{equation}
\label{eq:difference_increase}
d_{t}^{''} = (d_{t}^{'} - d_{t-1}^{'}) + (d_{t}^{'} - d_{t+1}^{'})
\end{equation}
It shows accuracy improvement in some noisy cases, such as reducing the error rate on the testing data from $2.27\%$ to $1.25\%$, with a 0.5 second interval. However, it will delay the decision for 1 additional time interval, since it requires the next feature $d_{t+1}$ in computation. 

\section{Conclusion and Future Work}
\label{sec:conclusion}

In this work, a noval real-time SCD approach using improved features through a speaker classification network is presented. The features are represented by vectors of attributes of the in-domain speakers, i.e. projected onto a space spanned by the in-domain speakers. It enables the use of simple distance metrics such as Euclidean distance between the feature centroids to detect speaker change in adjacent intervals. Using TIMIT data of 200 male speakers, the classifier guarantees to achieve 100\% accuracy, with speech no longer than 1 second. In the 15-minute synthesized conversations of 63 different speakers (62 unique speaker changes), theoretically there is only around 2\% of the changes are mis-detected with the F1 score above 0.98. It outperforms the results in  \cite{kotti2006automatic}, which also used TIMIT synthesized data, based on algorithms in \cite{lu2002speaker}. These results are still very competitive, compared to other algorithms using real world conversations \cite{ajmera2004robust,bigot2010exploiting}.

The next step is to test the algorithm with real-world conversations, where the number of speakers should be fewer and the speaker changes may less frequent, but they can be less predictable and speaker conflicts may occur. Since the Bayesian threshold depends on the speaker change frequency, which is unpredictable in real world scenarios, more robust and dynamic thresholding might be necessary to improve the performance. Second, better SCD performance has been observed with conversations from in-domain speakers. Thus, speaker clustering based on initial detection results is also desirable for converting new speakers into in-domain speakers and forming a better speaker classifier for feature transform. Third, currently the speaker classifier is trained to maximize classification accuracy for the in-domain speakers, rather than trained towards the best feature transformer for detecting speaker changes. Finally, how to select the speaker as in-domain speakers to train it for that purpose is still unclear and needs to be explored in the future.

\bibliographystyle{IEEEbib}
\bibliography{references}

\end{document}